\documentclass[%
superscriptaddress,
 amsmath,amssymb,
 aps,
prl,
twocolumn,
]{revtex4-2}
\usepackage[LGR,T1]{fontenc}
\usepackage[latin9]{inputenc}
\usepackage{float}

\usepackage{textcomp}
\usepackage{amssymb}
\usepackage{graphicx}
\usepackage{hyperref}
\usepackage{bibunits}

\usepackage{subscript}
\usepackage{amsmath}
\usepackage{color}
\usepackage{ulem}
\makeatletter
\DeclareFontEncoding{LGR}{}{}
\DeclareTextSymbol{\~}{LGR}{126}
\makeatother
\usepackage{babel}

\newcommand{\Tc}{T_{\rm c}}
\newcommand{\Tcr}{T_{{\rm c},\rho}}
\newcommand{\Tcc}{T_{{\rm c},ac}}

\newcommand{\UTe}{UTe$_{\rm 2}$}

\begin{document} 
\title{Thermodynamic and electrical transport properties of UTe$\rm_2$ under uniaxial stress}

\author{Cl\'ement Girod}
	\email{cgirod@lanl.gov}
	\affiliation{ Los Alamos National Laboratory, Los Alamos, New Mexico 87545, U.S.A.}
\author{Callum R. Stevens}
	\affiliation{School of Physics and Astronomy, University of Edinburgh, Edinburgh, U.K.}
\author{Andrew Huxley}
	\affiliation{School of Physics and Astronomy, University of Edinburgh, Edinburgh, U.K.}
\author{Eric D. Bauer}
	\affiliation{ Los Alamos National Laboratory, Los Alamos, New Mexico 87545, U.S.A.}
\author{Frederico B. Santos}
\author{Joe D. Thompson}
	\affiliation{ Los Alamos National Laboratory, Los Alamos, New Mexico 87545, U.S.A.}
\author{Rafael M. Fernandes}
	\affiliation{School of Physics and Astronomy, University of Minnesota, Minneapolis, Minnesota 55455, U.S.A}
\author{Jian-Xin Zhu}
	\affiliation{ Los Alamos National Laboratory, Los Alamos, New Mexico 87545, U.S.A.}
\author{Filip Ronning}
	\affiliation{ Los Alamos National Laboratory, Los Alamos, New Mexico 87545, U.S.A.}
\author{Priscila F. S. Rosa}
	\affiliation{ Los Alamos National Laboratory, Los Alamos, New Mexico 87545, U.S.A.}
\author{Sean M. Thomas}
	\email{smthomas@lanl.gov}
	\affiliation{ Los Alamos National Laboratory, Los Alamos, New Mexico 87545, U.S.A.}
	
\date{\today}

\begin{abstract}
Despite intense experimental efforts, the nature of the unconventional superconducting order parameter of \UTe~remains elusive. This puzzle stems from different reported numbers of superconducting transitions at ambient pressure, as well as a complex pressure-temperature phase diagram
. To bring new insights into the superconducting properties of \UTe, we measured the heat capacity and electrical resistivity of single crystals under compressive uniaxial stress $\sigma$ applied along different crystallographic directions. We find that the critical temperature
 $\Tc$ of the single observed bulk superconducting transition decreases with $\sigma$ along [100] and [110] but increases with $\sigma$ along  [001]. Aside from its effect on $\Tc$, we notice that 
  $c$-axis stress leads to a significant piezoresistivity, 
 which we associate with the shift of the zero-pressure resistivity peak at $T^* \approx 15 \, \rm K$ to lower temperatures under stress.
 Finally, we show that an in-plane shear stress $\sigma_{xy}$
 does not induce any observable splitting of the superconducting transition over a stress range of $\sigma_{xy}\approx0.17\,\rm GPa$. This result suggests that the superconducting order parameter of \UTe~may be single-component at ambient pressure.

\end{abstract}

\maketitle

 
The recently discovered uranium-based unconventional superconductor  \UTe~has attracted a lot of attention as a promising candidate for spin-triplet pairing and topological superconductivity \cite{ran2019}. Yet, despite intense experimental efforts (see Ref.~\cite{aoki2022}), the nature of the superconducting order parameter (OP) of \UTe~remains elusive. 
Among its numerous peculiarities, there is growing evidence that \UTe~hosts different superconducting phases as a function of applied hydrostatic pressure  \cite{braithwaite2019,thomas2020,aoki2020,ran2020,ran2021} and magnetic field \cite{ran2019_2,knebel2019,aoki2020,ran2020,ran2021}. However, the number of superconducting transitions at ambient pressure and zero magnetic field seems to depend on sample details and is thus still debated \cite{aoki2022}.

At zero field and ambient pressure, several studies suggested that \UTe~is a chiral superconductor based on the observation of chiral surface states \cite{jiao2020,bae2021}, a gap structure with point nodes \cite{metz2019}, and a broken time reversal symmetry (TRS) in the superconducting state \cite{hayes2021,wei2022}.
 Due to the presence of two distinct thermodynamic superconducting anomalies in some samples \cite{hayes2021}, \UTe~is proposed to possess two superconducting OPs. Additionally, the trainability of the polar Kerr signal with magnetic field along the crystallographic $c$ axis suggests that the product of these two superconducting OPs transforms as the $B_{1g}$ irreducible representation (irreps) of the orthorhombic point group D$_{2h}$ \cite{hayes2021,wei2022}. Importantly, because this point group has no multi-dimensional irreps, a description in terms of a two-component gap necessarily requires near-degenerate superconducting instabilities.

While some samples display two features in the specific heat across the superconducting transition, \UTe~crystals that display an optimal superconducting transition temperature $\Tc = 2\, \rm K$ and large residual resistivity ratios host a single thermodynamic superconducting transition, as manifested by a single jump in the specific heat \cite{cairns2020,rosa2021}.  
On the one hand, this could be an indication that the two OPs condense at very close temperatures that cannot be resolved in the specific heat. On the other hand, there has been no report of broken TRS in the superconducting phase of these samples showing a single transition. As a result, the nature of the superconducting OP is still unclear.
%
%

 In samples with a single superconducting transition (i.e. one jump in the specific heat), pressure splits it into two thermodynamic transitions that have opposite pressure dependence above $0.3\, \rm GPa$ \cite{braithwaite2019}. For samples with two peaks in the specific heat at ambient pressure, four peaks are observed above $0.3\, \rm GPa$ \cite{thomas2020}.
Moreover, upon further increasing the pressure, an antiferromagnetic phase emerges \cite{braithwaite2019,thomas2020}, which contrasts with the ferromagnetic fluctuations expected at ambient pressure \cite{sundar2019,ran2019}. The connection between this pressure-induced change in magnetic fluctuations and the pressure-induced splitting of the superconducting transitions remains unclear \cite{ishizuka2021}.

 \begin{figure*}[t]
\centering
\includegraphics[width=1\textwidth]{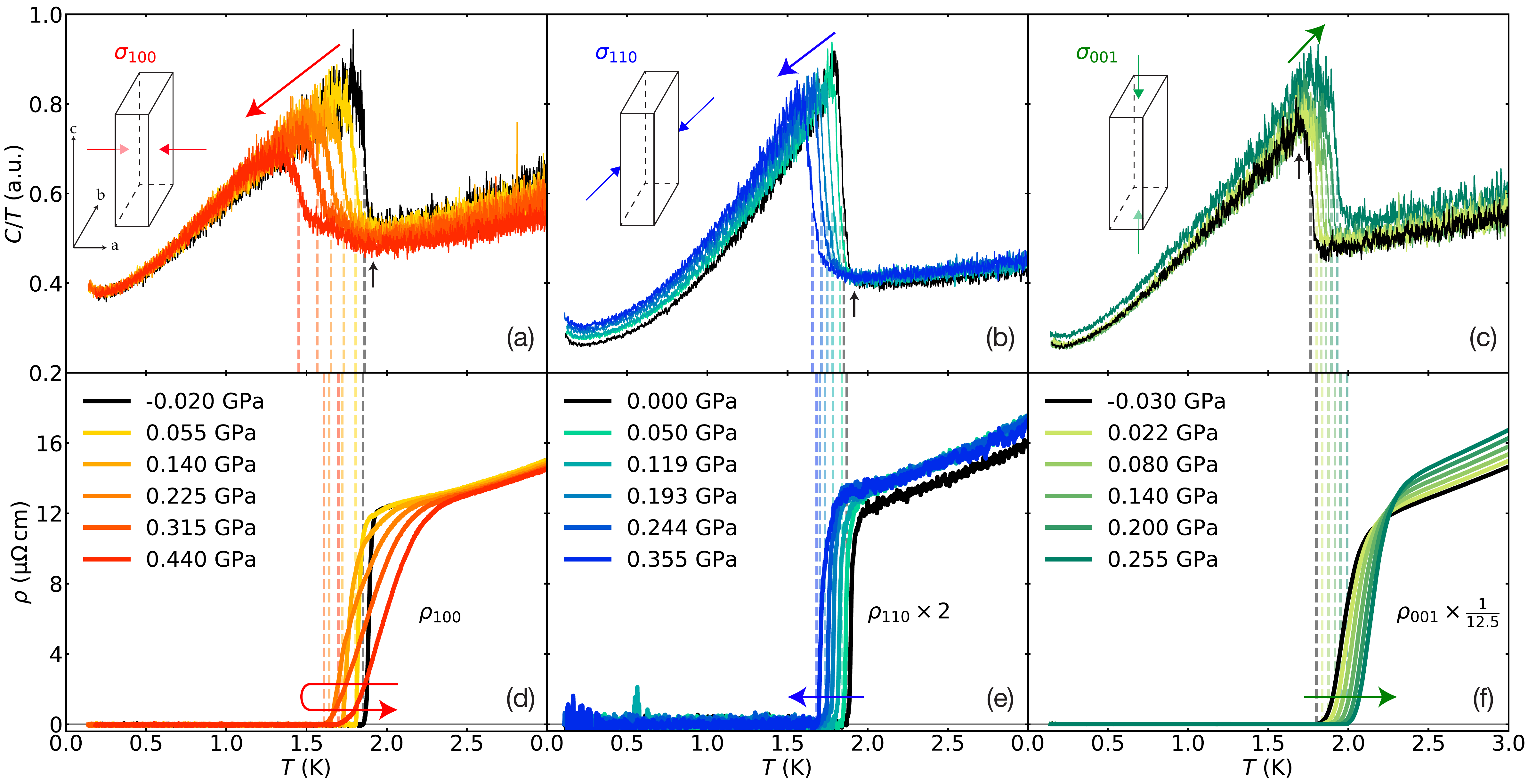}
\caption{Temperature dependence of the heat capacity $C/T$ (top panels) and electrical resistivity (bottom panels) along the indicated crystal directions ($\rho_{100}$, $\rho_{110}$ and $\rho_{001}$) at the indicated uniaxial stress values and orientations [(a), (d): $\sigma_{100}$; (b), (e): $\sigma_{110}$; (c), (f): $\sigma_{001}$]. ac calorimetry data were recorded with an excitation frequency $f\approx 20\, \rm Hz$. Colored dashed lines mark, for each stress value, the average temperature of the sharp rise of the jump in $C/T$ at $\Tcc$ on the top panels and the temperature $\Tcr$ below which the resistivity of the sample is zero on the bottom panels.  Colored arrows show the trends of $\Tcc(\sigma)$ and $\Tcr(\sigma)$ upon increasing the compressive stress. Black arrows on the top panels show the onset [(a), (b)] and end [(c)] of the heat capacity jump of the lowest stress curve (black lines). The sketches on the top panels show the direction of the applied stress. Resistivity values $\rho_{110}$ and $\rho_{001}$ are scaled ($\times 2$ and $\times 1/12.5$ respectively) for clarity.  The difference between the resistivity curves at the two lowest $\sigma_{110}$ values is due to a crack in the sample, not to reversible piezoresistivity.}
\label{Fig1}
\end{figure*}

  
Uniaxial stress has proved to be a powerful tool to study multi-component superconductors. This technique has been used extensively to study the phase diagram and the OP of $\rm Sr_2RuO_4$  \cite{li2020,grinenko2021}. While this material was initially thought to display spin-triplet pairing \cite{ishida1998}, recent NMR data performed under strain demonstrated it to be actually a singlet superconductor \cite{pustogow2019}.

Here we investigate the nature of the superconducting OP of \UTe~by measuring the low temperature ac heat capacity and the electrical resistivity of single crystals under uniaxial stress $\sigma_{100}$, $\sigma_{110}$ and $\sigma_{001}$  respectively applied along the [100], [110] and [001] crystallographic directions (details in the Methods section of Supplemental Materials). Our unstressed samples display a single transition according to the specific heat.
 We report that the $\Tc$ value extracted from calorimetry decreases with compressive $\sigma_{100}$ and $\sigma_{110}$ but increases with compressive $\sigma_{001}$.
   Most importantly, we show that a symmetry-breaking in-plane shear stress $\sigma_{xy}$ does not induce any observable splitting of the superconducting transition. We discuss different scenarios to explain these results, including one in which the superconducting OP of \UTe~is single-component and another one in which the two components belong to different symmetry channels from those proposed previously \cite{hayes2021,wei2022}.
  Aside from its effect on $\Tc$, $c$-axis stress induces a significant piezoresistivity, presumably caused by
 the reduction of the energy scale corresponding to the feature observed at $T^\star\approx 15\, \rm K$, which has been attributed to the of onset short-range magnetic correlations or anisotropic Kondo coherence \cite{willa2021,kang2021,eo2021}.


\begin{figure*}[t]
\centering
\includegraphics[width=1\textwidth]{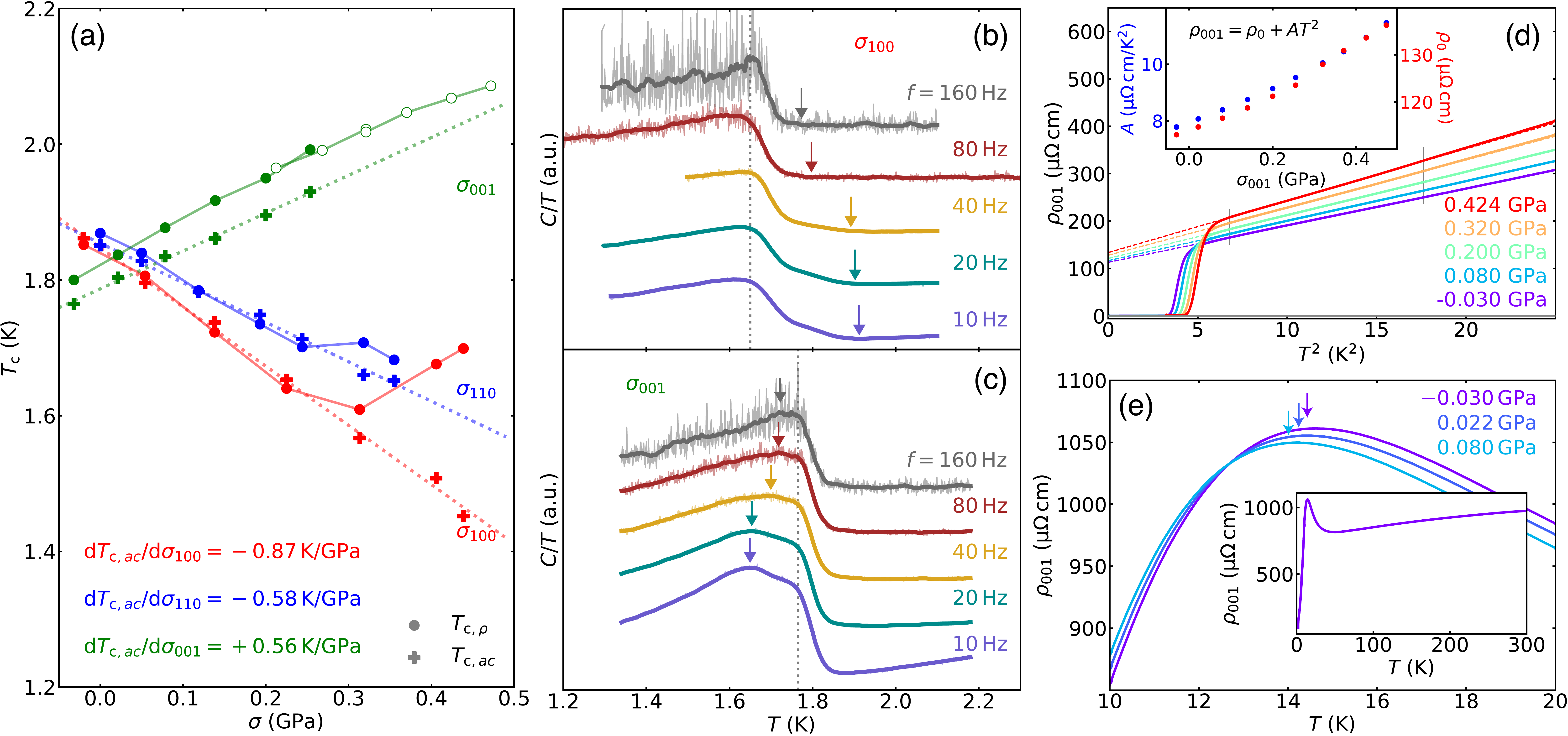}
\caption{\textbf{(a)} Uniaxial stress dependence of the superconducting transition temperatures  $\Tcr$ (circles) and $\Tcc$ (crosses) obtained for applied  $\sigma_{100}$ (red),  $\sigma_{110}$ (blue) and $\sigma_{001}$ (green) directions. Full symbols correspond to data from Fig.~\ref{Fig1} and open symbols from panel (d) at higher $\sigma_{001}$ values. Colored dashed lines are linear fits to $\Tcc(\sigma)$, whose slopes $\frac{{\rm d}\Tcc}{{\rm d}\sigma}$ are displayed on the lower left corner of the figure. 
\textbf{(b) \& (c)} Temperature dependences of the heat capacity $C/T$ under compressive stress $\sigma_{100}=0.050\,\rm GPa$ [panel (b)] and $\sigma_{001}=0.200\,\rm GPa$ [panel (c)], at the indicated oscillating current heater frequencies. 5-8 times higher powers were used to improve resolution and raw data (light colors) are smoothed (dark colors) to better show the different features. For clarity, amplitudes have been renormalized and offset to show a comparable jump amplitude. Dashed lines indicate the temperature of the main heat capacity jump, that shows no frequency dependence. Colored arrows show the evolution of the onset temperature of the foot above $\Tc$~[panel (b)] and shoulder below $\Tc$~[panel (c)].
 \textbf{(d) \& (e)} Temperature dependence of the $c$-axis resistivity $\rho_{001}$ at different $c$-axis stress levels $\sigma_{001}$.  (d) Resistivity below $5\, \rm K$ and plotted versus $T^2$ together with $\rho_{001} = \rho_0 +AT^2$ fits to the data between $2.6\, \rm K$ and $4.2\, \rm K$ (dashed lines). The inset shows the $\sigma_{001}$ dependance of the $\rho_0$ and $A$ parameters extracted from the fits. (e) Stress evolution of the position of the peak at $T^\star \approx 15\, \rm K$ (colored arrows). The inset shows the data at $\sigma_{001}=-0.030\, \rm GPa$ below room temperature. }
\label{Fig2}
\end{figure*}

The ac calorimetry data displayed in Figs.~\ref{Fig1}~(a)~-~(c) show the temperature dependence of the heat capacity plotted as $C/T$ of three samples at the indicated $\sigma_{100}$, $\sigma_{110}$ and $\sigma_{001}$ uniaxial stress values. Note that in our convention positive $\sigma$ means compressive strain, whereas negative means tensile. The curves at the lowest stress show a single and sharp transition at the thermodynamic superconducting critical temperature $\Tcc$, defined as the average temperature of the sharp rise in $C/T$ that occurs when most of the sample becomes superconducting (dashed vertical lines). A single superconducting transition is in agreement with the characterization data of the unstressed samples (S.M.~Fig.~2) and with the results of Refs.~\cite{cairns2020,rosa2021}. 

Figures~\ref{Fig1}~(d)~-~(f) show the temperature evolution of the electrical resistivity $\rho_{100}$, $\rho_{110}$ and $\rho_{001}$ with current along the applied stress direction. These measurements were carried out simultaneously with ac calorimetry. For all  stress and current directions, the resistivity at the lowest stress value displays a sharp transition to the superconducting state at a resistive critical temperature $\Tcr$ (below which $\rho=0$) that is in good agreement with the one extracted from heat capacity. The small difference between them (less than $0.1\, \rm K$) is easily explained by different definitions of $\Tcc$ and $\Tcr$. 

We observe two main effects upon application of compressive uniaxial stress. First,  $\Tcc$ changes monotonically upon applying stress along all directions. For $\sigma_{100}$ and $\sigma_{110}$, $\Tcc$ decreases with increasing stress, while $\Tcc$ increases with increasing $\sigma_{001}$. The evolution of  $\Tcc$ and $\Tcr$ with $\sigma_{100}$, $\sigma_{110}$ and $\sigma_{001}$ is summarized in the stress versus temperature phase diagram shown in Fig.~\ref{Fig2}~(a). As expected for bulk superconductivity, the stress evolution of $\Tcr$ tracks that of $\Tcc$ for most stress values; the difference between $\Tcc$ and $\Tcr$ for $\sigma_{100}>0.250\, \rm GPa$ will be discussed later.
%

Using Erhenfest's relation, we find that the opposite trend of $\Tcc$ with $\sigma_{100}$ (and $\sigma_{110}$) as compared to $\sigma_{001}$ is in agreement with a previous thermal expansion study \cite{thomas2021} that reported a negative jump at $\Tc$ in the linear thermal expansion coefficient along $[100]$ (and $[010]$), but a positive jump for the coefficient along the $[001]$ direction.
These results suggest that the superconducting state observed at ambient pressure is favored by a smaller $c$-axis length  and larger $a$-axis length. 
Because a U-U dimer (shortest U-U distance) in the crystal structure of \UTe~is along the $c$ axis, and uranium chains run along the $a$ axis, our results support previous theoretical arguments that the $c$-axis dimer is key to the formation of the superconducting state in \UTe~\cite{shishidou2021,miao2020}.

The second effect of applied uniaxial stress is the appearance of a foot above $\Tcc$ for $\sigma_{100}$ and $\sigma_{110}$ and a shoulder below  $\Tcc$~for $\sigma_{001}$ [see Figs.~\ref{Fig1}~(a)~-~(b) and Fig.~\ref{Fig1}~(c), respectively]. As the number of  superconducting transitions in \UTe~remains a central question \cite{rosa2021,hayes2021}, the origin of these features must be understood. It is clear from Figs.~\ref{Fig1}~(a)~-~(b) that the foot above $\Tcc$ for all $\sigma_{100}$ and $\sigma_{110}$  values coincides with the onset temperature of the superconducting anomaly of the lowest stress curve. For $\sigma_{001}$, the shoulder for all stress values extends above the temperature of the $C/T$ jump at $\Tcc$ of the lowest stress curve, as shown in Fig.~\ref{Fig1}~(c). The most likely explanation for these features is thus an inhomogeneous stress in the samples, especially in regions of the samples that extend underneath the mounting plates of the stress cell. 

To test
 the possibility of inhomogeneous stress, we measured heat capacity at different excitation frequencies $f$. As  $f$ increases, ac calorimetry probes an increasingly smaller, more homogeneously stressed volume of the sample \cite{li2020}. 
   The results for $\sigma_{100}=0.050\, \rm GPa$ and $\sigma_{001}=0.200\, \rm GPa$, which are the stress directions that cause the largest effects, are respectively shown in Figs.~\ref{Fig2}~(b) and (c). At the lowest excitation frequency ($f=10\,\rm Hz$), the foot and shoulder features are clearly visible.
    As frequency increases, the position of the main heat capacity jump at $\Tcc$ remains essentially unchanged. In comparison, the shoulder and foot onset temperatures seem to progressively merge with $\Tcc$, as expected when probing a more homogeneously stressed part of the sample.
 This shows that stress inhomogeneity is responsible for these two artifacts
  and that there is a \textit{single} heat capacity jump related to bulk superconductivity at the indicated stress levels. 
%

Our main finding 
is the absence of splitting of the superconducting transition for applied stress along [100], [110] and [001], as seen in Figs.~\ref{Fig1}~(a)~-~(c). 
This is especially meaningful for $\sigma_{110}$ which, unlike $\sigma_{100}$ and $\sigma_{001}$, breaks the orthorhombic symmetry of the crystal to monoclinic by lowering the point group symmetry from D$_{2h}$ to C$_{2h}$. By expressing the stress tensor of $\sigma_{110}$ into the basis of the main crystallographic axes, one finds $\sigma_{110}=0.68\sigma_{100}+0.32\sigma_{010}+0.47\sigma_{xy}$, where $\sigma_{xy}$ is the in-plane shear stress and $\sigma_{010}$ stress along the [010] direction. 
Because the resulting shear strain $\epsilon_{xy}$ transforms as the $B_{1g}$ irrep of D$_{2h}$,  it is expected to split the transition temperatures of the two proposed nearly-degenerate OPs \cite{hayes2021}, since their product also transforms as $B_{1g}$. The difference between the two nearly-degenerate superconducting transition temperatures $\Delta T_{\rm c}$ is expected to follow (to leading order in the applied strain):
\begin{equation*}
\Delta T_{\rm c}= \sqrt{\Delta T_{{\rm c}( \epsilon_{xy}=0)}^{2}+\lambda^2 \epsilon_{xy}^2}, 
\end{equation*}
where $\Delta T_{{\rm c}( \epsilon_{xy}=0)}$ is the unstressed splitting of the superconducting transition temperatures  ($\Delta T_{{\rm c}( \epsilon_{xy}=0)}\approx0$  in our case, since a single transition is observed in the unstressed samples) and $\lambda$ is a coupling constant (see S. M. section E for details of the calculations). 

The fact that we do not observe any noticeable splitting of the superconducting transition upon application of $\sigma_{110}$ [see Fig.~\ref{Fig1}~(b)] may
be explained by different scenarios.  One possibility is that the OP is different from the one proposed previously, either because there is only a single superconducting OP or because the nearly-degenerate OPs belong to symmetry channels that would not allow for a coupling that is linear in $\epsilon_{xy}$. Another possible explanation for the absence of detectable splitting of the superconducting transition induced by $\sigma_{110}$ would be a $\lambda$ value too small to cause any appreciable splitting for $\sigma_{110}<0.355\, \rm GPa$.
 Using the elastic tensor obtained from density functional theory (DFT) detailed in S.M. section D,  we find that at that stress level, the induced $\epsilon _{xy}\approx -0.6\%$ would not lead to a $\Delta T_{\rm c}$ greater than $0.1 \, \rm K$ for $\lambda<16\, \rm K$). In this regard, we note that, as detailed in Ref.~\cite{rosa2021}, the presence of two superconducting anomalies in some unstrained samples could be related to a lower sample quality.  

If  \UTe~hosts a single superconducting OP, regardless of its symmetry, one expects $\Tc$ to evolve quadratically with shear strain $T_{\rm c}= T_{\rm c}^{(\epsilon_{xy}=0)}+\lambda \epsilon_{xy}^2$. From Fig.~\ref{Fig2}~(a), we see that $\Tcc$ evolves linearly with $\sigma_{110}$, like $\sigma_{100}$ and $\sigma_{001}$, but with a different slope. This implies that the $\sigma_{110}$ response is dominated by the symmetry-preserving stress along the main axes, $\sigma_{100}$ and $\sigma_{010}$, rather than the symmetry-breaking stress $\sigma_{xy}$. Symmetry considerations imply that $\Tc$ should show a linear dependence on $\sigma_{100}$ and $\sigma_{010}$.


  From Fig.~\ref{Fig2}~(a), we determine $\frac{{\rm d} \Tcc }{ {\rm d} \sigma_{100}}\approx-0.87\,\rm K/GPa$ and $\frac{{\rm d} \Tcc }{ {\rm d} \sigma_{001}}\approx+0.56\,\rm K/GPa$. 
Their sum gives $\sim-0.31\,\rm K/GPa$, which
 upon comparison with both prior  hydrostatic pressure studies \cite{braithwaite2019,thomas2020} ($\frac{{\rm d} \Tc }{ {\rm d} P}\approx-0.5\,\rm K/GPa$) and thermal expansion and specific heat ($\frac{{\rm d} \Tc }{ {\rm d} P}\approx-0.49\,\rm K/GPa$) using Erhenfest's relation \cite{thomas2021} suggests that the evolution of $\Tc$ under applied $\sigma_{010}$ is smaller than that of the two other axes. 
 Using the DFT elastic tensor shown in S.M. section D, we find that the evolution of $\Tc$ with $\sigma_{100}$ and $\sigma_{001}$ cannot be explained in terms of strain along a single direction through Poisson expansion. This suggests that there is no dominant strain direction controlling $\Tc$ in \UTe. 
It would be interesting to apply higher $c$-axis stress, which seems to be tuning the system in a different direction than hydrostatic pressure \cite{braithwaite2019,thomas2020}.  

Under hydrostatic pressure, the superconducting transition splits into two above $0.3\, \rm GPa$, leading to a slight slope change of the lower $\Tc (P)$ and an initial enhancement of the higher $\Tc (P)$. The latter undergoes a drastic suppression for $P>1.2\, \rm GPa $ \cite{braithwaite2019,thomas2020} due to the emergence of a magnetic ground state. Here, the absence of splitting of the superconducting transition and the linear evolution of  $\Tcc$ with uniaxial stress approaching $0.3\, \rm GPa$ ($\sigma_{001}$) or exceeding this value ($\sigma_{100}$ and $\sigma_{110}$) suggests that higher stress levels would be required to drive the system to a regime with a different ground state. 
However, for $\sigma_{100}>0.250\, \rm GPa$, $\Tcr$ starts to increase with increasing stress, in contrast to the behavior of $\Tcc$ [see Fig.~\ref{Fig2}~(a)]. In addition, the resistive superconducting transition shows substantial broadening upon application of $\sigma_{100}$ whereas its width barely increases with $\sigma_{110}$ and $\sigma_{001}$, as shown in Figs.~\ref{Fig1}~(d)~-~(f). This behavior was verified in another sample with applied $\sigma_{100}$ (see S.M.~Fig.~3).
 
  For hydrostatic pressures just above $0.3\, \rm GPa$, the emerging superconducting transition that splits from the main $\Tc(P)$ curve displays a positive ${\rm d} \Tc / {\rm d} P$ as well as both an initially small signature in heat capacity and a significant broadening in resistivity \cite{braithwaite2019,thomas2020}. A similar scenario could then take place under applied  $\sigma_{100}$, based on the analogous resistive behavior. In this case, a stress value of $\sigma_{100}\approx 0.3 \, \rm GPa$ could be just enough to drive the system towards the regime in which a split superconducting transition emerges. This would mean that shorter U-U distance in the chains along the $a$ axis could be a key ingredient for the enhancement of the second superconducting phase observed at high pressures. Alternatively, this effect may also be caused by the presence of filamentary or surface superconductivity. The application of higher stress would be useful
 to distinguish between the two scenarios, since only in the first one the higher-temperature transition observed in resistivity would be manifested in the heat capacity.

Finally, we turn to the pronounced piezoresistivity  observed above $\Tcr$ for stress and current along the $c$ axis [see Fig.~\ref{Fig1}~(f)]. This effect was not observed for stress applied along the other crystal directions [Figs.~\ref{Fig1}~(d)~and~(e)].
By fitting the normal-state $\rho_{001}$ over the extended temperature range shown in Fig.~\ref{Fig2}~(d) to $\rho_{001} = \rho_0 +AT^2$, we observe an 
enhancement, upon increasing compressive $\sigma_{001}$, of both the coefficient associated with electron-electron scattering $A$ and the residual resistivity $\rho_0$.  
 The observed enhancement of $A$ over a stress range of about $0.5\, \rm GPa$ is non-trivial ($30 \%$)
  but smaller that the approximately factor of two increase in $A$ with applied hydrostatic pressure of $0.56\, \rm GPa$ \cite{braithwaite2019} and much smaller than the $\sim 1000\%$ increase in $A$ at a metamagnetic transition near $ 32\, \rm T$ for magnetic fields applied close to the $b$ axis at atmospheric pressure \cite{knafo2021}.
   These comparisons suggest that at $\sigma_{001}\approx0.5\, \rm GPa$, the system is still away from any possible quantum critical point. 
Additionally, we find that the peak in $\rho_{001}$ around $T^ \star\approx15\, \rm K$ that was previously reported in Ref.~\cite{eo2021}, shifts towards lower temperatures as $\sigma_{001}$ increases,
  as displayed in Fig.~\ref{Fig2}~(e). This peak has been attributed to either a Kondo energy scale \cite{kang2021} or
  to the onset of short-range magnetic correlations \cite{willa2021}. The large low temperature $c$-axis piezoresistivity could therefore be caused by the decrease of the energy scale associated with the peak in $\rho_{001}$ at $T^ \star$ with $\sigma_{001}$.


In conclusion, our combined low temperature ac calorimetry and electrical transport measurements under uniaxial stress show that the single thermodynamic superconducting transition at $\Tcc$ in \UTe~has opposite evolution under applied compressive stress $ \sigma_{100},\sigma_{110}$ ($\frac{{\rm d}\Tcc}{{\rm d}\sigma_{100}},\frac{{\rm d}\Tcc}{{\rm d}\sigma_{110}}<0$) and  $ \sigma_{001}$ ($\frac{{\rm d}\Tcc}{{\rm d}\sigma_{001}}>0$). This result is consistent with conclusions from previous thermal expansion studies and suggests that superconductivity in \UTe~is favored by a smaller $c$-axis length and larger $a$-axis length. 
Additionally, we observe a relatively large piezeoresistivity above $\Tc$ for current and stress applied along the $c$-axis, suggesting that aside from enhancing superconductivity, $\sigma_{001}$ is responsible for a decrease in a different and yet unidentified normal-state energy scale \cite{willa2021,kang2021,eo2021}. Applying higher $ \sigma_{001}$ values could be interesting to determine if the system can be tuned towards a different ground state.
 For $\sigma_{100}>0.25 \, \rm GPa$, we observe a difference between the  $\Tcc$ and $\Tcr$ values, similar to what is observed under hydrostatic pressure exceeding $0.3 \, \rm GPa$, where the signatures of a second superconducting phase become unambiguous \cite{braithwaite2019,thomas2020}. 
Our results suggest that a stress level $\sigma_{100}\approx 0.25 \, \rm GPa$ could be just enough to drive the system towards this regime.
%
Finally, through the application of a symmetry breaking shear stress, $\sigma_{xy}$, we fail to observe clear signatures for two split superconducting transitions, as one would have expected for the case of two nearly-degenerate superconducting OPs whose product transforms as $B_{1g}$. This implies either that the coupling between shear strain and the superconducting OPs is very small or that the superconducting OP of \UTe~is different from the one proposed previously. In the latter scenario, TRS breaking might be explained by the condensation of a sub-leading superconducting instability near dislocations and other lattice defects, similarly to what has been recently proposed to explain TRS breaking in Sr$_2$RuO$_4$ \cite{willa2021_2}. To disentangle these different scenarios, it would be interesting to perform Kerr effect measurements as in Ref.~\cite{hayes2021} on crystals showing a single superconducting transition.  


The development of ac calorimetry under uniaxial stress was supported by the Laboratory Directed Research and Development program. The remaining experimental work and crystal synthesis at Los Alamos were supported by the U.S. Department of Energy (DOE), Office of Basic Energy Sciences, Division of Materials Science and Engineering project "Quantum Fluctuations in Narrow Band Systems". JXZ (density functional theory calculations) was supported by Quantum Science Center, a U.S. DOE Office of Science National Quantum Information Science Research Center, and in part by Center for Integrated Nanotechnologies, a U.S. DOE Office of Basic Energies Science user facility, in partnership with the LANL Institutional Computing Program for computational resources. CS and AH (crystal synthesis) acknowledge support from UK EPSRC grant EP/P013686/1. RMF (phenomenological modeling) was supported by the U.S. DOE, Office of Science, Basic Energy Sciences, Materials Science and Engineering Division, under award no. DE-SC0020045.

\bibliographystyle{apsrev4-2} 
\bibliography{Main}
%

%

\end{document}


\cleardoublepage
\newpage

\section*{Supplementary Material}

\subsection{Methods}

\UTe~single crystals were grown using the chemical vapor transport method described in references~\cite{rosa2021,cairns2020}. Samples used in this study display a single superconducting transition in heat capacity  with $\Tc\approx 1.8\, \rm K$ (see S.M. Fig.~\ref{UTe2_Cp_comparison}). Bar shaped samples with typical dimensions $\rm 2\, mm \times 0.5 \, mm \times 0.15\, mm$ were prepared, with longest dimensions along the [100], [110] and [001] crystallographic directions using a Laue diffractometer (two different [100] samples and one for each of the other directions were studied,  as shown on S.M. Fig.~\ref{SM_Fig3}). Samples properties are shown in S.M. Table.~\ref{table1}.

 \begin{table}[H]
\caption{Properties of the studied samples: name, unstressed critical temperature $\Tc$ from heat capacity measurement (from S.M. ~Fig.~\ref{UTe2_Cp_comparison}), sample dimensions, inside and outside distance between the electrical contacts, gap between the stress cell clamps and slope of $\Tcc(\sigma)$ for the associated uniaxial stress direction.}
\begin{ruledtabular}
\begin{tabular}{cccccc}
Sample name & Unstressed $T_{\rm c}\,$(K) & $l\times w\times t\, (\rm \mu m^3)$ &Contact distance $(\rm \mu m)$ & Gap $(\rm \mu m)$ & $\frac{{\rm d}\Tcc}{{\rm d}\sigma}\, (\rm K/GPa)$ \\
\hline
100 & $1.85$ & $2307\times462\times170$ &$248-547$& 1168  & -0.87 \\
100 B & 1.66 & $1800\times380\times149$ &$235-615$ &1181 & -0.88 \\
\hline
110 & 1.87 & $3000\times624\times150$ &$264-588$& 2013 & -0.58 \\
\hline
001 & 1.77& $1400\times303\times195$ &$172-404$& 850 & +0.56 \\
\end{tabular} 
\end{ruledtabular}
\label{table1}
\end{table}

\begin{figure}[H]
\centering
\includegraphics[width=0.4\textwidth]{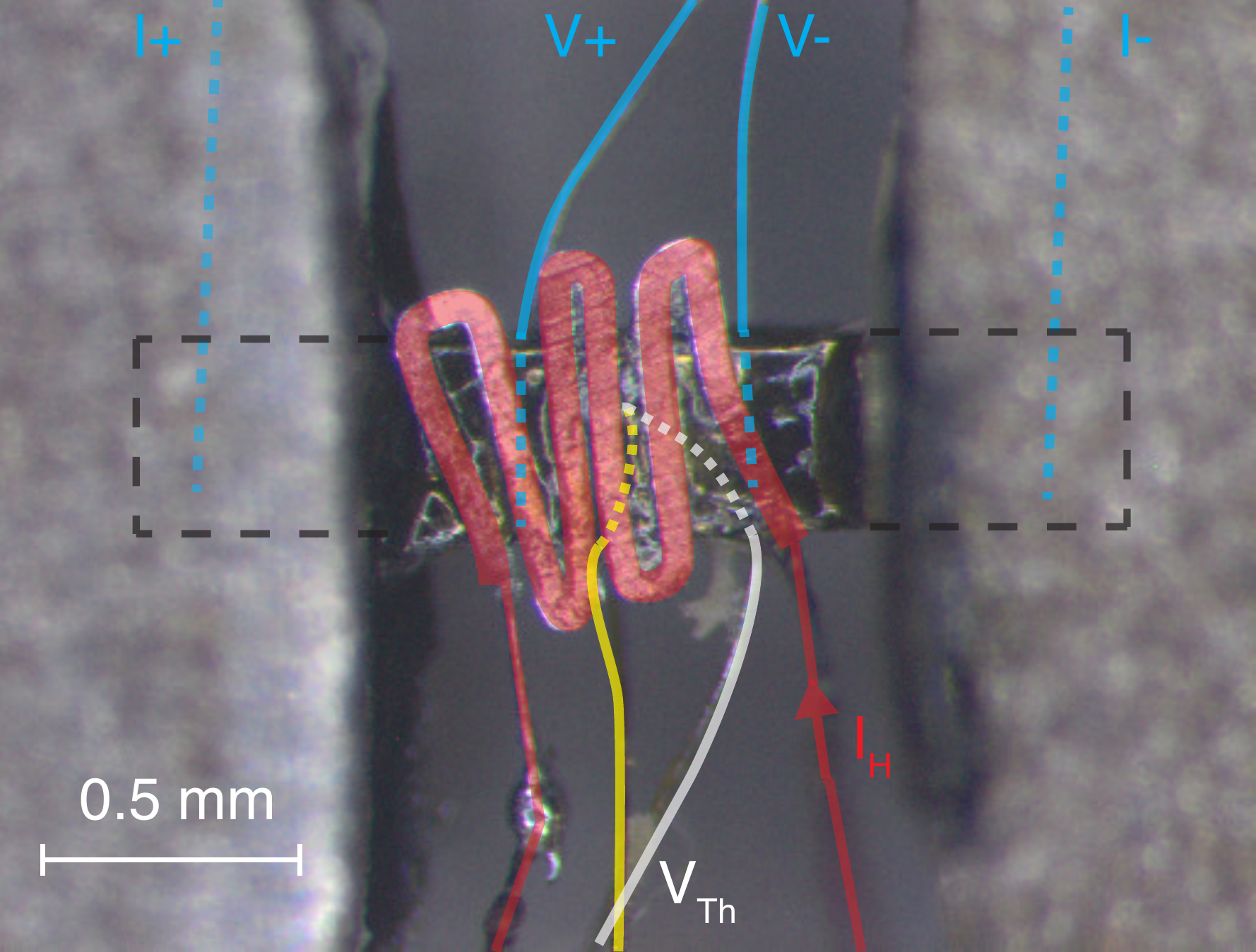}\caption{Photograph of the experimental setup showing the sample (in black and black dashed lines) held with epoxy between the cell clamps (grey). Solid and dashed blue lines indicate the location of the leads for resistivity measurement. Red lined show the constantan heater and its current leads deposited on top of the sample using GE varnish. Yellow and white lines respectively show the Au/Fe(0.07\%) and chromel thermoucouple wires connected underneath the sample using GE varnish.}
\label{experimental_setup}
\end{figure}

\begin{figure}[H]
\centering
\includegraphics[width=0.9\textwidth]{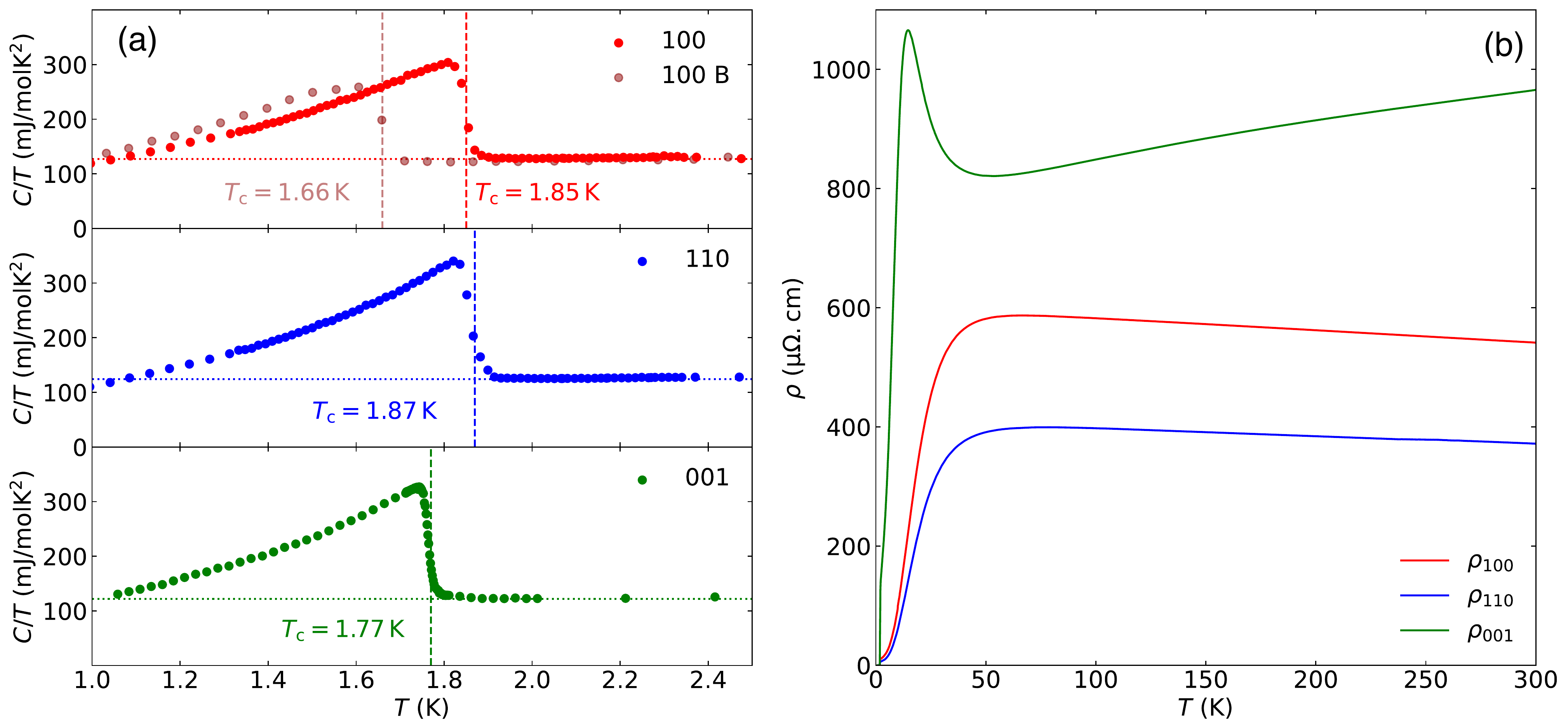}
\caption{ \textbf{(a)} Specific heat data of the three samples showed in the paper and of the additional 100 B sample measured before the samples were mounted in the stress cell. Dashed lines indicated the $\Tc$ values defined as the average temperature of the superconducting jump. Dotted lines represent the normal state specific heat. \textbf{(b)} Temperature dependence of the resistivity $\rho_{100}$, $\rho_{110}$ and $\rho_{001}$ of our unstressed samples 100, 110 and 001.}
\label{UTe2_Cp_comparison}
\end{figure}

\begin{figure}[H]
\centering
\includegraphics[width=0.9\textwidth]{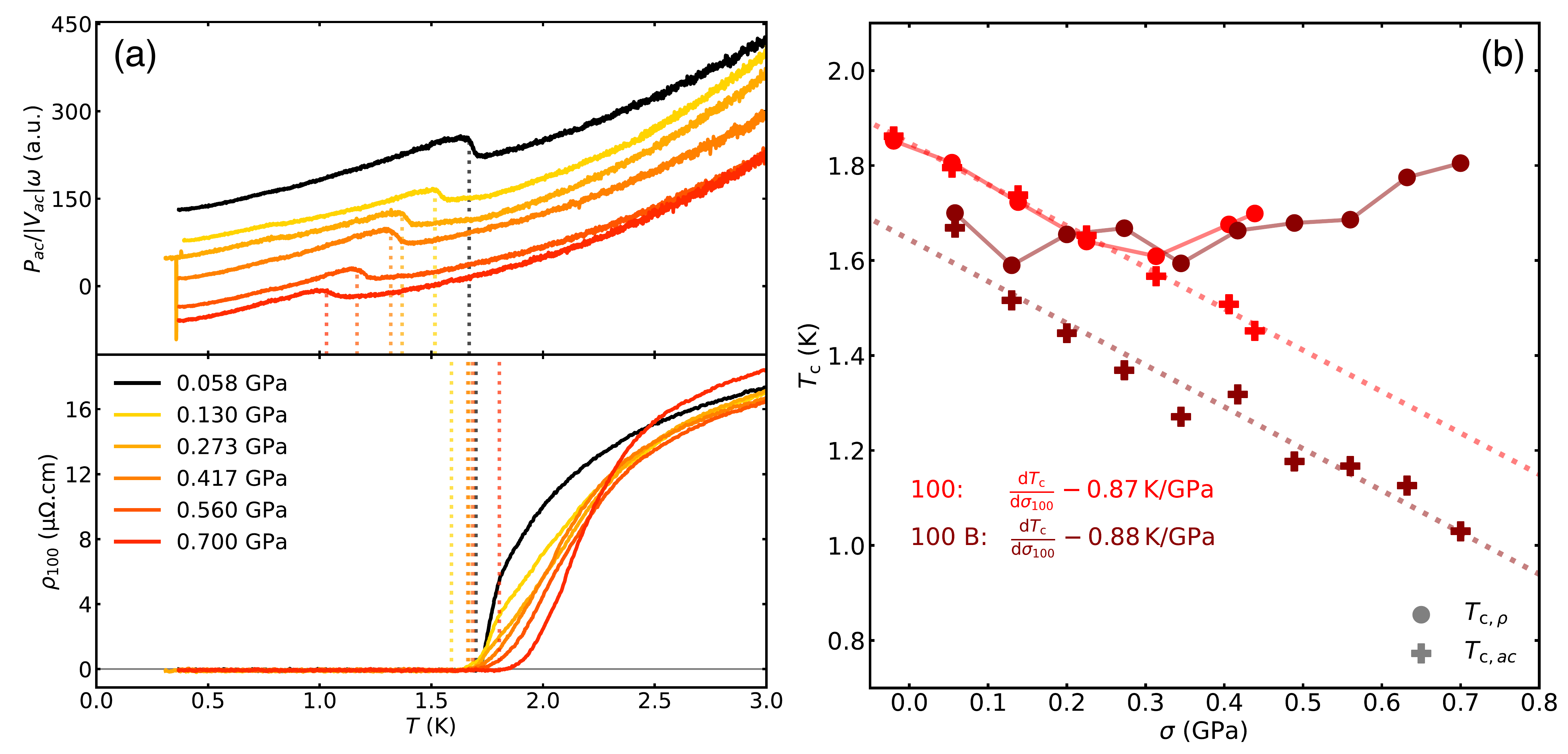}
\caption{\textbf{(a)} Temperature dependence of the heat capacity $C/T$ (top panel) and electrical resistivity $\rho_{100}$ (bottom panel) of sample 100 B at the indicated  $\sigma_{001}$ values. Colored dashed lines mark the average temperature of the sharp rise of the jump in $C/T$ at $\Tcc$  and the temperature $\Tcr$ below which the resistivity of the sample is zero.
 \textbf{(b)} Stress dependence of $\Tcr$ (circles) and $\Tcc$ (crosses) for our samples 100 (red) and 100 B (maroon). Circles represent $\Tcr$ and crosses $\Tcc$ extracted from a). Colored dashed lines are linear fits $\Tcc(\sigma)$, which slopes $\frac{{\rm d}\Tcc}{{\rm d}\sigma}$ are displayed on the lower left corner of the figure.}
\label{SM_Fig3}
\end{figure}

Uniaxial stress was applied along the length of the samples using a commercial stress cell (Razorbill FC100), resulting in stresses $\sigma_{100}=\sigma_{xx}$, $\sigma_{110}\approx 0.68 \sigma_{xx}+ 0.32 \sigma_{yy}+0.47 \sigma_{xy}$ and $\sigma_{001}=\sigma_{zz}$.  $\sigma_{110}$  was decomposed in the basis of the main axes by rotating the $\sigma_{110}$ stress tensor by an angle $\theta=\arctan{a/b} \approx 34.16^\circ$ in the $ab$ plane. Stress levels reaching $\sigma\approx 0.3-0.4\, \rm GPa$ could be obtained without damaging the samples. Samples were glued between the clamps of the stress cell using Stycast 2850FT epoxy with catalyst 9. Electrical contact to the sample was made using platinum wires and silver paint deposited on gold pads sputtered directly onto the samples. Electrical resistance was measured using a Lakeshore 372 resistance bridge. Temperature oscillations for the AC heat capacity measurements were generated using a constantan wire heater ($R\approx \rm 15 \, \Omega$) using powers of order $P_{AC}\approx \rm 1-10 \, \mu W$ at frequencies of $2f\approx \rm 10-320 \, Hz$.  The induced temperature oscillations $T_{AC}$ were recorded using a chromel-Au/Fe(0.07\%) thermocouple and a SR860 lockin amplifier after being amplified  by a SR554 low-noise transformer preamplifier. Both the heater and thermocouple were mechanically and thermally connected to the samples using GE varnish. The experimental setup is shown in S.M. Fig.~\ref{experimental_setup}. In this configuration, the heat capacity of the system is related to $C\propto P_{AC}/\left\vert T_{AC}\right \vert f$. All measurements were carried out either in a CMR adiabatic demagnetization refrigerator, allowing measurements down to $0.15\, \rm K$, or in a Quantum Design PPMS for sample characterization and measurements above $3\, \rm K$.

The thermodynamic critical temperature $\Tcc$ of the samples was determined as the average temperature of the sharp rise of $C/T$ occurring when most of the sample becomes superconducting. This sharp rise can be distinguished from the smoother foot above $\Tcc$ observed for applied $\sigma_{100}$ and $\sigma_{110}$.

Applied stress level was determined using the capacitance versus force calibration curve of the stress cell. Force was then converted into stress by dividing it by the cross section of the sample perpendicular to the applied stress direction. The dimensions of the samples, shown in S.M. Table.~\ref{table1}, were measured using an optical microscope, with a typical resolution of $10\, \mu m $.
%
%
%
%
%

As stress is either increased or decreased, we noticed slight changes in the thermal coupling between the heater/thermocouple and the samples. This results in a (small) decrease in the temperature oscillations (with the same excitation current) on measurement carried out consecutively at different stress levels. This effect explains the small increase of the normal state value of $C/T$ with stress observed in Main Fig.~1~(a), for which data were recorded with decreasing $\sigma_{100}$, and the opposite effect for Main Figs.~1~(b)~-~(c), for which data were recorded with increasing $\sigma_{110}$ and $\sigma_{001}$.

\subsection{Unstressed characterization}

The residual resistivity above $\Tcr$, $\rho_0$, has values of the order of $10\, \rm \mu \Omega\, cm$ for $\rho_{100}$ and $\rho_{110}$, and $100\, \rm \mu \Omega\, cm$ for $\rho_{001}$. These values are consistent with previous reports of the resistivity of \UTe~along the different crystal directions \cite{ran2019,eo2021}
.  Although we do not find that $\rho_{100}< \rho_{110} $, this is probably due to a fairly large uncertainty of the geometric factors of our samples. In order to measure resistivity over a more homogeneously stressed region of the sample, the voltage leads are centered in the gap between the mounting plates and placed close together ($150-600\, \rm \mu m$).

\subsection{Additional 110 sample}

\subsection{DFT calculation of the elastic constants}
The elastic constants were calculated by using density functional theory (DFT) with the pseudopotential projector-augmented wave method \cite{SM__Kresse1999} implemented in the Vienna Ab initio Simulation Package (VASP) \cite{SM__Kresse1996,SM__Kresse1993}.  A local density approximation to the exchange-correlation functional combined with the correlation correction \cite{SM_Liechtenstein1995} with $U=6 \, \rm eV$ and $J=0.57 \, \rm eV$ to the U-5f orbitals was used. The spin-orbit coupling was incorporated. In the nonmagnetic constraint, the simulations give the fully relaxed lattice constants $a = 4.11\,  \angstrom$, $b = 6.13\,  \angstrom$, $c = 13.93\,  \angstrom$, in good agreement with the experimental measurements at ambient conditions \cite{SM_Boehme1992}. Further relaxation calculations were then performed for varying uniaxial strains applied along $a$- and $c$-axis. For each structure, the elastic moduli were then evaluated. The obtained DFT elastic tensor (in GPa) is:
 
 \begin{equation*}
 \begin{bmatrix}
\sigma_{xx} \\
\sigma_{yy} \\
\sigma_{zz} \\
\sigma_{xy} \\
\sigma_{yz} \\
\sigma_{zx} \\
\end{bmatrix}
=
 \begin{bmatrix}
96.96782    & 40.94466 &48.58602     & 0 & 0 & 0    \\
40.94466       & 140.74513     &46.67250           & 0 & 0 & 0    \\
48.58602           & 48.58602         &101.32160                & 0 & 0 & 0    \\
0           & 0         &0                & 27.06222      & 0 & 0    \\
0           & 0         &0                & 0      & 19.61040      & 0    \\
0           & 0         &0                & 0      & 0 & 57.25525    \\
\end{bmatrix}
 \begin{bmatrix}
\epsilon_{xx} \\
\epsilon_{yy} \\
\epsilon_{zz} \\
\epsilon_{xy} \\
\epsilon_{yz} \\
\epsilon_{zx} \\
\end{bmatrix}.
\end{equation*}
  
\subsection{Phenomenological model for the effect of shear strain}

Consider two nearly degenerate superconducting instabilities, with
gaps $\Delta_{\gamma_{1}}$ and $\Delta_{\gamma_{2}}$, that transform
as different one-dimensional irreducible representations $\gamma_{1}$ and $\gamma_{2}$ of the D$_{2h}$ point group.
The Kerr data \cite{hayes2021} indicates that the quantity $i\left(\Delta_{\gamma_{1}}\Delta_{\gamma_{2}}^{*}-\Delta_{\gamma_{1}}^{*}\Delta_{\gamma_{2}}\right)$
has the same symmetry as an out-of-plane magnetic field $H_{z}$,
which transforms as $B_{1g}$ and is odd under time-reversal. As discussed
in Ref. \cite{hayes2021}, this restricts $\gamma_{1}$ and $\gamma_{2}$
to the following combinations: $\left(B_{3u},B_{2u}\right)$, $\left(B_{1u},A_{u}\right)$,
$\left(B_{3g},B_{2g}\right)$, $\left(B_{1g},A_{g}\right)$. Thus,
for any of these combinations, the time-reversal even quantity $\left(\Delta_{\gamma_{1}}\Delta_{\gamma_{2}}^{*}+\Delta_{\gamma_{1}}^{*}\Delta_{\gamma_{2}}\right)$
transforms as $B_{1g}$, i.e. as shear strain $\epsilon_{xy}$. To
quadratic order in $\Delta_{\gamma_{i}}$ and to leading order in
$\epsilon_{xy}$, the Landau free-energy expansion is:

\begin{equation}
F=a_{1}\left|\Delta_{\gamma_{1}}\right|^{2}+a_{2}\left|\Delta_{\gamma_{2}}\right|^{2}+\tilde{\lambda}\epsilon_{xy}\left(\Delta_{\gamma_{1}}\Delta_{\gamma_{2}}^{*}+\Delta_{\gamma_{1}}^{*}\Delta_{\gamma_{2}}\right)
\end{equation}

Here, $\tilde{\lambda}$ is a coupling constant, $a_{i}=a_{i,0}\left(T-T_{c,i}^{(0)}\right)$
with $a_{i,0}>0$ and $i=1,2$, and $T_{c,i}^{(0)}$ are the unstressed
transition temperatures associated with the two superconducting instabilities.
After minimizing the relative phase between the two order parameters,
we can rewrite this expression in the matrix form:

\begin{equation}
F=\left(\begin{array}{cc}
\left|\Delta_{\gamma_{1}}\right| & \left|\Delta_{\gamma_{2}}\right|\end{array}\right)\left(\begin{array}{cc}
a_{1} & -\left|\tilde{\lambda}\epsilon_{xy}\right|\\
-\left|\tilde{\lambda}\epsilon_{xy}\right| & a_{2}
\end{array}\right)\left(\begin{array}{c}
\left|\Delta_{\gamma_{1}}\right|\\
\left|\Delta_{\gamma_{2}}\right|
\end{array}\right)
\end{equation}

Diagonalization of the matrix gives two superconducting transition
temperatures given by the condition $a_{1}a_{2}=\left|\tilde{\lambda}\epsilon_{xy}\right|$.
We find:

\begin{align}
T_{c,1} & =\left(\frac{T_{c,1}^{(0)}+T_{c,2}^{(0)}}{2}\right)+\sqrt{\left(\frac{T_{c,1}^{(0)}-T_{c,2}^{(0)}}{2}\right)^{2}+\frac{\lambda^{2}\epsilon_{xy}^{2}}{4}}\\
T_{c,2} & =\left(\frac{T_{c,1}^{(0)}+T_{c,2}^{(0)}}{2}\right)-\sqrt{\left(\frac{T_{c,1}^{(0)}-T_{c,2}^{(0)}}{2}\right)^{2}+\frac{\lambda^{2}\epsilon_{xy}^{2}}{4}}
\end{align}
where we defined $\lambda\equiv2\tilde{\lambda}/\sqrt{a_{1,0}a_{2,0}}$.
Thus, the splitting $\Delta T_{c}$ between the two transitions depends
on $\epsilon_{xy}$ according to:

\begin{equation}
\Delta T_{c}=\sqrt{\left(\Delta T_{c}^{(0)}\right)^{2}+\lambda^{2}\epsilon_{xy}^{2}}
\end{equation}
where $\Delta T_{c}^{(0)}$ is the splitting between the two transitions
in the absence of applied stress.

\bibliographystyle{apsrev4-2} 
\bibliography{SM}
%